\documentclass[journal]{IEEEtran}

\usepackage{graphicx}
\usepackage{placeins}
\usepackage{amsmath}
\usepackage[linesnumbered,ruled]{algorithm2e}

\ifCLASSINFOpdf
\else
\fi

\begin{document}
\title{High-Accuracy Inference in Neuromorphic Circuits
using Hardware-Aware Training}

\author{Borna~Obradovic,
        Titash~Rakshit,
				Ryan~Hatcher,
				Jorge~A.~Kittl,
        and~Mark~S.~Rodder%} <-this % stops a space
\thanks{B. Obradovic, T. Rakshit, R. Hatcher, J.A. Kittl, and M.S. Rodder are with the 
Samsung Advanced Logic Lab, Austin TX 78754, USA.}% <-this % stops a space

\thanks{(e-mail: b.obradovic@samsung.com)”}% <-this % stops a space
\thanks{Manuscript received August 15th, 2018}

%\thanks{D. Lin, N. Waldron and N. Collaert are with, IMEC Leuven, Belgium.}% <-this % stops a space
}
\maketitle

\begin{abstract}
Neuromorphic Multiply-And-Accumulate (MAC) circuits utilizing synaptic weight elements based on 
SRAM or novel Non-Volatile Memories (NVMs)
provide a promising approach for highly efficient hardware representations of neural networks. 
NVM density and robustness requirements suggest that off-line training is the right choice for ``edge''
devices, since the requirements for synapse precision are much less stringent. 
However, off-line training using ideal mathematical weights and activations can result
in significant loss of inference accuracy when applied to non-ideal hardware. Non-idealities
such as multi-bit quantization of weights and activations, non-linearity of weights, finite
max/min ratios of NVM elements, and asymmetry of positive and negative weight components all result in degraded inference accuracy. In this work, it is demonstrated
that non-ideal Multi-Layer Perceptron (MLP) architectures using low bitwidth 
weights and activations
can be trained with negligible loss of inference accuracy
relative to their Floating Point-trained counterparts using a proposed off-line, continuously differentiable HW-aware 
training algorithm. The proposed algorithm is applicable to a wide range of hardware models, and
uses only standard neural network training methods. The algorithm is demonstrated on the MNIST and EMNIST
datasets, using standard MLPs.

\end{abstract}


% \fi
\IEEEpeerreviewmaketitle

\section{Introduction}
\IEEEPARstart{H}{ardware} accelerators for Deep Neural Nets (DNNs) based on neuromorphic approaches
such as analog resistive crossbar arrays are receiving significant attention due to their potential
to significantly increase computational efficiency relative to standard CMOS approaches. An important
subset are accelerators designed for inference only, i.e. utilizing off-line training \cite{Strukov1, jeds}. The absence of on-chip
training capability results in simplified, smaller area weight implementations, as well as a reduced complexity
of the peripheral circuitry. The converse case of on-chip training requires precise weights and activations (at least 6-bit precision) due to the small weight increments required by the gradient descent (and related) algorithms
\cite{SYu2, SYu3, GBurr1, GBurr2}
While there are several possible approaches to implementing high weight precision \cite{SYu3, burr_nature}, all of them incur an area or
programming variability penalty relative to simple, low-precision weights that are sufficient for the inference-only case. In this paper, the assumption is made that near-term applications for neuromorphic accelerators on mobile SoCs will not benefit from on-chip training,
which is instead relegated to the cloud. The focus is on improved inference performance and power reduction.

The key disadvantage of the inference-only approach is that any discrepancy
of the on-chip weights and activations from the off-line ideals results in a potentially significant
loss of inference accuracy. The most obvious example is the quantization of weights and activations,
but also includes any non-linearities of the weights (w.r.t. input signal or target weight), as well 
as a finite max/min ratio and sign asymmetry. Applying a ``brute-force'' algorithm which maps off-line trained weights 
to HW elements can result in significant loss of inference capability. The solution to this problem
is the emulation of the behavior of HW during the off-line training process. Instead of training
a mathematically ``pure'' DNN, the traning is performed on a model of the DNN 
implementation in target hardware. This  process is referred to as ``Hardware-Aware Training''. The paper is organized as follows.
The description of example HW-architectures is presented in Sec. \ref{sec:arch}.
The general training algorithm suitable for a wide range of HW-architectures is 
presented in Sec. \ref{sec:training}.
Individual applications of the algorithm of Sec. \ref{sec:training} on the HW-architectures
of Sec. \ref{sec:arch} are shown on in Sec. \ref{sec:results}.

\FloatBarrier
\section{Hardware Architectures}
\label{sec:arch}
While many hardware architectures for MAC have been considered, three specific examples
are discussed in this work. All are based on the Ferroelectric FET (FeFET) NVM for weight storage
\cite{jeds}.
There is no particular significance to the choice of FeFET in the context of HW-aware training; it
is merely used here for the purpose of example. Other NVMs, or even SRAM would have resulted in similar considerations. 
Furthermore, the chosen examples are interesting in the context of this work because of the different training challenges that they
represent, not necessarily because they are the best choices for neuromorphic implementations.

\subsection{Binary XNOR}
\label{sec:xnor}

\begin{figure}[!ht]
\centering
\includegraphics[width=3.0in]{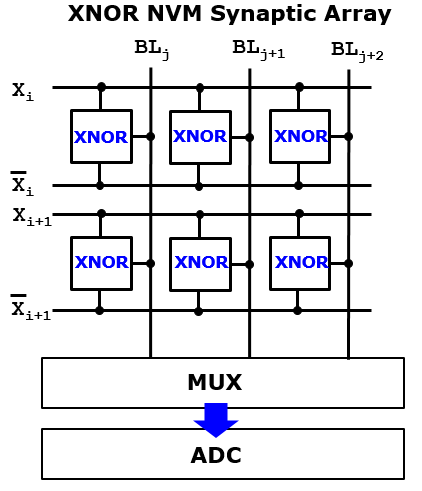}
% where an .eps filename suffix will be assumed under latex, 
% and a .pdf suffix will be assumed for pdflatex; or what has been declared
% via \DeclareGraphicsExtensions.
\caption{The XNOR array is shown. Each XNOR cell has two inputs ($X$ and $\bar{X}$), and 
an output that discharges a bitline (BL). Inference can take place in sequential, row-by-row fashion,
or with multiple rows at once. In the latter case, the degree of BL discharge is an analog quantity
(proportional to number of ``1s'' in XNOR cell outputs), and an ADC is required in the readout.}
\label{xnor_array}
\end{figure}

\begin{figure}[!ht]
\centering
\includegraphics[width=3.0in]{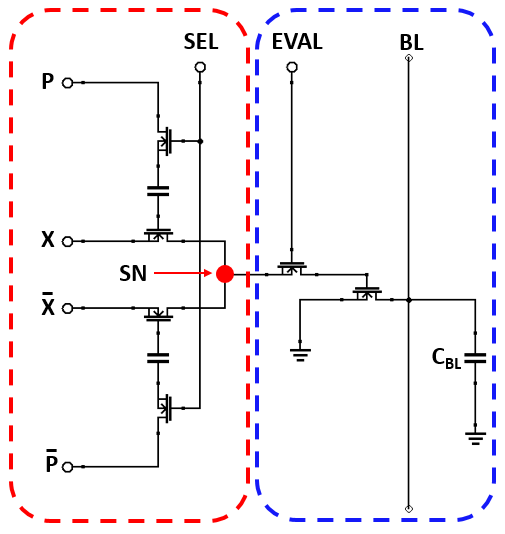}
% where an .eps filename suffix will be assumed under latex, 
% and a .pdf suffix will be assumed for pdflatex; or what has been declared
% via \DeclareGraphicsExtensions.
\caption{The XNOR cell implemented using dynamic FeFET logic is illustrated. The circuit
in the dashed red box implements the XNOR(X, W), storing the result as a dynamic voltage
on the node SN. Evaluation is performed by enabling EVAL and partially discharging the 
pre-charged bitline BL. Each XNOR cell which is activated at SN contributes to a partial
discharge of the bitline.}
\label{FeFET_XNOR}
\end{figure}

The first architecture considered is an XNOR \cite{xnornet} implementation using FeFET-based dynamic logic.
The array architecture is shown in Fig. \ref{xnor_array}, while the FeFET-based XNOR circuit block
is illustrated in Fig. \ref{FeFET_XNOR}. XNOR-based networks could also be realized in SRAM, with identical
training considerations. The FeFET-based approach is shown as an example here because other cell characteristics
(such as area) are more desirable than in the SRAM case.

The cell of Fig. \ref{FeFET_XNOR} implements the function $XNOR(X, W)$ where $X$ is a logic input, applied
to the sources of the FeFETs, while $W$ is the weight, stored as the state of polarization of the BEOL FeCaps
(connected to the gates of the underlying FETs of the overall FeFET). The FeFETs are programmed by applying
moderately high voltage pulses to the program lines. Write disturbs for XNOR cells which share program lines
are prevented using the selector FETs. Inference is performed by applying input signals ($X$, $\bar{X}$) with
the PRG inputs grounded. This causes the storage node SN to either charge up to VDD
or stay at GND. At the same time, the bitline BL is pre-charged. This performs
the binary multiply portion of the MAC operation. The accumulate portion is performed
in the second phase of the inference; the EVAL signal is enabled, and the driver FETs
perform a partial discharge of the bitline. 
This particular approach to MAC eliminates
(to a great extent) the variability problem associated with the FeCaps (which may be
significant for scaled FeCaps), since the final voltage of the SN is either $\approx$ V$_{DD}$
or $\approx$ 0. Variability is nevertheless present due to the Vt variation of the driver FETs,
leading to variability in the discharge rate of the BL. This scheme is therefore useful when
parallel accumulation is desired and FeFET variability is much greater than that of the 
standard  FETs.

\FloatBarrier
\subsection{Ternary Conductive Crossbar}
\label{sec:Crossbar}

\begin{figure}[!ht]
\centering
\includegraphics[width=3.0in]{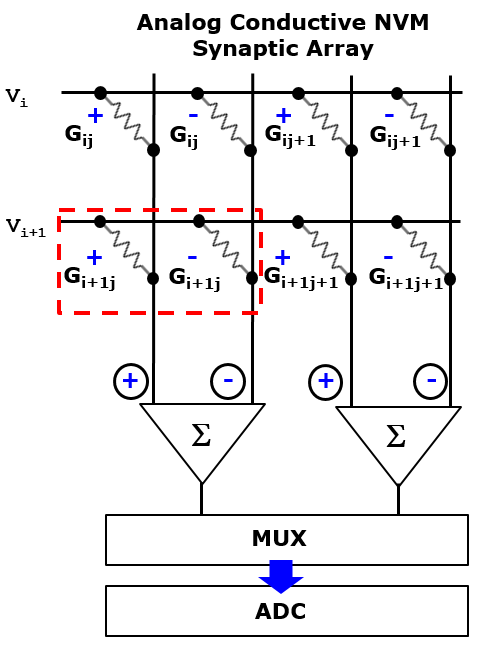}
% where an .eps filename suffix will be assumed under latex, 
% and a .pdf suffix will be assumed for pdflatex; or what has been declared
% via \DeclareGraphicsExtensions.
\caption{The cross-bar array utilized in this work is illustrated. The array consists of two sets
of weights: one each for positive and negative conductance contributions. The ``positive'' and ``negative''
weight cells are identical; the minus sign is introduced using a current mirror on the negative output line,
just prior to the summing amplifier. }
\label{analog_array}
\end{figure}

\begin{figure}[!ht]
\centering
\includegraphics[width=2.5in]{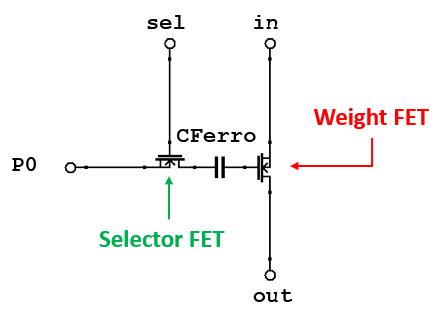}
% where an .eps filename suffix will be assumed under latex, 
% and a .pdf suffix will be assumed for pdflatex; or what has been declared
% via \DeclareGraphicsExtensions.
\caption{A single weight cell of the Ternary crossbar array is shown. The weight cell is based
on a FeFET (shown using a FET with a ferroelectric capacitor connected to the gate) and uses
a select FET attached to the gate of the weight FET in order to prevent write disturbs.
Programming is performed by applying high voltage pulses to P0 while $sel$ is high. During
inference, P0 is grounded and $sel$ is high. The conductive state of the weight FET is determined
by the stored polarization of the FeCap. }
\label{one_bit}
\end{figure}

The Ternary Conductive Crossbar (TCC) architecture is a slightly modified resistive cross-bar array \cite{GBurr1, GBurr2}, as shown in Fig. \ref{analog_array}. The individual weight cells are shown in Fig. \ref{one_bit}.
The standard approach of using two weights to represent positive and negative conductances is used. The modification
of the standard approach arises only in the use of dedicated program lines (Fig. \ref{one_bit}), one for each row of
weights. The program lines are shared across the entire row; write disturb prevention is accomplished
by activating individual select lines. In inference mode, the program lines are grounded, and the weights 
behave like two-terminal devices, forming a cross-bar between the signal input and output lines. The weights
themselves are FeFETs; the conductance level of the FeFETs (each with a grounded gate terminal) determines their
weight value. Details of the programming can be found in \cite{jeds}.

In principle, it is possible to store analog or multi-bit weights into a single FeFET, if 
sufficiently accurate programming is available, and FeCap variability is sufficiently controlled. 
In this work, however, the FeFETs are assumed to be purely
digital devices, programmed into either a strongly ON-state or strongly OFF-state. Such a scheme
avoids the complexities of multi-bit or analog programming, as well as the challenges of variability
when defining multiple programming states. Since the $G^+/G^-$ conductance pair is utilized, three
useful  conductance states are available: ($G_{ON}$, $G_{OFF}$),  ($G_{OFF}$, $G_{ON}$),
and  ($G_{OFF}$, $G_{OFF}$), corresponding to the mathematical states {1, -1, 0}. Since the
FeFETs themselves are acting as the weight elements, all FeFET non-idealities impact  the weight
as well. While programming is similar to that of the XNOR cell, inference must be performed
with low signal voltages which ensure that the FeFETs are in the linear regime throughout the 
inference event. Larger voltages create non-linear distortions which must be taken into account
in the network model. Due to the presence of the $G^+/G^-$ pair, the nominal value of the ternary
``zero'' is exactly zero (due to $G^+/G^-$ cancellation). Statistically, however, there will be 
some variability of the ternary zero due to process variability. Additionally, any imperfections
of the current mirror used to obtain the $G^-$ conductance will result in $G^+, G^-$ asymmetry.

\subsection{Multi-Bit Crossbar}
\begin{figure}[!ht]
\centering
\includegraphics[width=2.5in]{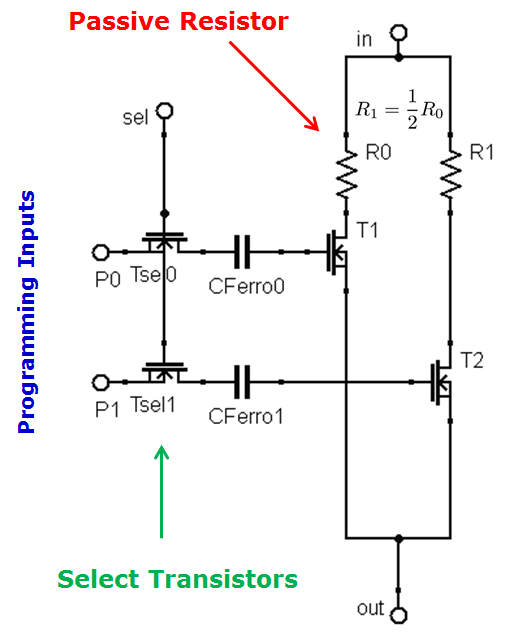}
% where an .eps filename suffix will be assumed under latex, 
% and a .pdf suffix will be assumed for pdflatex; or what has been declared
% via \DeclareGraphicsExtensions.
\caption{The schematic of a two-bit weight cell is illustrated. The resistance is provided by 
passive resistors, while the FeFET transistors enable or disable the individual branches. A select
transistor is provided for each bit. Additional bits require additional parallel branches
of resistor-transistor combinations. During inference, the program inputs are grounded
and the weight cell acts as a two-terminal device. The values of the passive resistors are
chosen to provide a binary ladder of overall weight conductance.}
\label{two-bit}
\end{figure}

A possible approach to multi-bit crossbars (two bits in this case) is illustrated in 
Fig. \ref{two-bit} \cite{jeds}.
The weight cell consists of multiple branches, each comprised of a series combination of an FeFET
and a passive resistor. The conductance values of the resistors follow a binary ladder. The total
conductance of the weight is then given by:
\begin{equation}
G_{tot} = \sum_{k=0}^{n-1} b_k 2^k G_0^k
\end{equation}

\noindent where $n$ is the total number of branches (i.e. bits) available ($n$=2 
for the purposes of this work). The overall crossbar array program and inference
operation are identical to those of the standard FeFET crossbar, as described in
Sec. \ref{sec:Crossbar}. However, while the latter implemented a Ternary weight, the
circuit described in this section implements a Quinary weight: a signed two-bit weight, for a total
of five possible weight levels. As in the case of the single FeFET case, the behavior of the
weight is subject to the non-idealities of the FeFET and the passive resistor. However,
this structure is far more tolerant to FeFET non-linearity, since the conductance is primarily
dependent on that of the passive resistors. In cases where the FeFET conductance (in the ON state)
is not multiple orders of magnitude higher than that of the passive resistor, a non-linear
skew of weight values must be taken into account during training for best inference accuracy.

\section{Hardware-Aware Training}
\label{sec:training}

As shown in previous work \cite{jeds, SYu2, SYu3}, direct Binary, Ternary, and even 
Quinary quantization of weights described in the
previous section results in degraded inference accuracy relative to Floating-Point (FP) implementations. This accuracy
loss is further compounded by non-idealities of the weights. The key problem, however, is not
a fundamental lack of inference capability of the non-ideal networks. Instead, it is the mismatch
between the training model and the inference model used. This is not necessarily a priori obvious, 
but is demonstrated
in this work and elsewhere \cite{xnornet, gxnornet}.
The standard training procedure generally assumes
that the full range of real numbers (or at least their FP representation) is available for weights and activations.
Training is typically performed using a suitably chosen variant of Gradient Descent with Backpropagation used to
compute the gradient. The set of weights thus obtained is then quantized, using a quantization spacing
that optimizes inference accuracy on a validation set. Binary and Ternary quantizations thus obtained show a significant accuracy loss \cite{jeds}; Quinary quantizations are more acceptable, but nevertheless show some accuracy degradation relative to the pure SW case \cite{jeds}. Of equal importance are deviations from ideality that are not related to quantization, such as asymmetries and non-linearities in the weight HW. In order to properly take into account non-idealities of specific HW implementations, a HW-aware training
algorithm is proposed.

\subsection{Training Algorithm}

The proposed training algorithm takes into account the appropriate HW model in a way which permits
discontinuities, such as those caused by quantization, without sacrificing continuous differentiability
of the cost function. All other non-idealities of the weights and activations are likewise included.
As summarized in Alg. \ref{algorithm}, the HW-aware training algorithm starts with a pre-trained neural net, resulting from FP-based training. Any suitable training method can be used to obtain 
$w_{FP}$. Regularization may or may not be performed for $w_{FP}$ at this stage. As discussed in \cite{jeds}, 
regularization without the appropriate HW model results in a loss of generalization capability, so re-training
with HW-aware regularization is necessary.
Next, the algorithm proceeds through several iterations of HW-aware weight refinement. At each iteration, the 
weight matrices are replaced by HW-description functions $g_{HW}(w, X, \alpha)$. The g-functions
represent the non-ideal, HW-induced weights, and are therefore functions of the ``ideal'' weights, 
as well as the input vector $X$. Additionally, the g-functions depend on the variable $\alpha$, which
serves as a continuation parameter. With $\alpha=0$, $g(w, X, \alpha)=w$; with $\alpha=1$, 
$g(w, X, \alpha)=G_{HW}(w, X)$, i.e. an accurate representation of the HW. With any intermediate
value of the $\alpha$ parameter, the g-functions represent an approximation of the true HW model,
becoming increasingly accurate as $\alpha$ is increased. A similar approach is used for activation functions, though
these are generally functions of the pre-activation value $z$ only. Hyperparameter optimization forms
the outer loop of the algorithm. Hyperparameters include standard parameters for $L^1$,
or $L^2$ regularization, but also parameters related to the $g_{HW}$ functions. The latter include parameters
such as the quantization step $\Delta$, but not parameters related to the physical model of the circuit
such as asymmetries and non-linearities. In general, hyperparameters should only include variables
which control the training, not those that characterize the hardware. It should also be noted that the
network model used for validation should always contain the exact (or ``best'') model of the HW, not the
HW model used for the current training iteration. This ensures that regularization is being used in a context
most similar to the final test evaluation.

\begin{algorithm}[!ht]
    \SetKwInOut{Input}{Input}
    \SetKwInOut{Output}{Output}

    \underline{HWApproxTrain} 
    \Input{Training set $(x_{train},\ y_{train})$, validation set $(x_{valid},\ y_{valid})$, 
		       hardware-approximation function $g_{HW}(w, X, \alpha)$, 
		       neural network model $nn_{model}$, sequence of HW-approximation parameters $\{\alpha\}$, 
					weights from FP-based training $w_{FP}$}
    \Output{$w_{optim}$, the trained weights}
		
		\textbf{Set} $w_{iter} = w_{FP};$ \\
		
		\For{$\lambda^k$ in $\{Hyperparams\}$} { 
			\For{$\alpha_i$ in $\{\alpha_{init}, ...\ \alpha_{final}\}$}{
			
				 $nn_{iter}(w,\ X)=nn_{model}(g_{HW}(w,\ X,\ \alpha_i), X);$ \\
			
				 $w_{iter}^k$ = $ADAM(nn_{iter},\ w_{iter}^k, \lambda^k, x_{train}, y_{train});$ \\
				}
				$Cost^k = CrossEntropy(y_{valid}, nn(w_{final}, x_{valid}))$ 
			}
			$opt=argmin(Cost^k)$ \\	
		\Return{$w_{iter}^{opt}$;}
    \caption{Hardware-Aware Training Algorithm}
		\label{algorithm}
\end{algorithm}

\begin{figure}[!ht]
\centering
\includegraphics[width=3.5in]{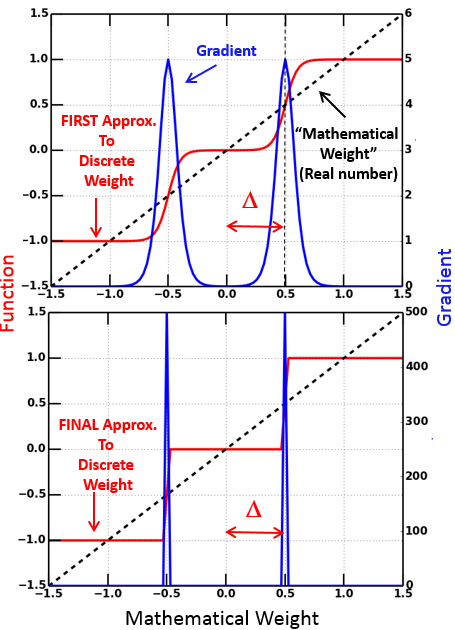}
% where an .eps filename suffix will be assumed under latex, 
% and a .pdf suffix will be assumed for pdflatex; or what has been declared
% via \DeclareGraphicsExtensions.
\caption{The $g^3_{HW}$ weight HW-model is shown in two separate iterations. The top figure
illustrates the first iteration in which the HW-model is applied, where the transition between weight levels is gradual.
The bottom figure represents the HW-model at the end of the training iterations. Transitions
are essentially abrupt. Gradients are large only very near weight transitions; in the case of
the near-ultimate HW-model, very few weights actually get modified by gradient descent, and
the algorithm comes to a stop.}
\label{weight_approx}
\end{figure}

A key property of the g-functions is that they are continuously differentiable
w.r.t. $w$ for all values of $\alpha$, except possibly for $\alpha=1$. This ensures that the exact
gradient is available at all steps in the iteration, enabling gradient descent-type optimizers
(such as the suggested ADAM) to perform efficiently. The weights obtained in each iteration are used as 
the initial guess in the next iteration. After a few iterations of the algorithm (often just one),
the obtained weights $w_{iter}$ are optimized for an accurate representation of the HW model.
The algorithm can be illustrated with a simple example: a neural network model with undistorted
Ternary weights. Discrete weight levels of the Ternary system can be approximated
as follows:

\begin{equation}
g^3_{HW} = 2 \Delta \bigg[ \sigma 
\bigg(\frac{w-\Delta}{w_{sc}}\bigg)
+
\sigma \bigg(\frac{w+\Delta}{w_{sc}} \bigg) 
- 1
\bigg]
\label{eq:weights}
\end{equation}

\noindent where the superscript $3$ is the number of weight categories of the weights (3 for Ternary),
$w$ represents the mathematical (FP) weights, $\Delta$ is the level spacing parameter
(for uniform spacing, as used in this example), and $w_{sc}$ is the weight-level transition scale
parameter. The latter sets the variable scale of the sigmoid ($\sigma$) of Eqn. \ref{eq:weights}.
The scaled sigmoid provides a smooth step function, and the
conversion from mathematical (FP) weights to discrete weight levels is then accomplished in a 
continuously differentiable fashion by adding an appropriately scaled smooth step function 
(as in Eqn. \ref{eq:weights}.)
As the weight scale parameter $w_{sc}$ is reduced, the steps become increasingly abrupt, 
asymptotically approximating discrete hardware. Thus, the $w_{sc}$ parameter is a proxy for the $\alpha$
``hardware approximation parameter'' of Alg. \ref{algorithm} (and can be directly related to it in a 
number of possible ways, for example $w_{sc}=1-\alpha$). For any given $w_{sc}$ (or equivalently $\alpha$)
the hardware approximation function $g^n_{HW}$ of Eqn. \ref{eq:weights} is continuously differentiable,
enabling the use of standard gradient-driven optimizers with backpropagation. Two steps in the
iterative refinement algorithm are illustrated in Fig. \ref{weight_approx}.

\begin{figure}[!ht]
\centering
\includegraphics[width=3.75in]{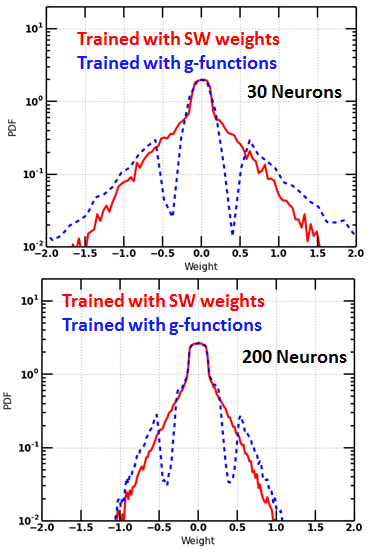}
% where an .eps filename suffix will be assumed under latex, 
% and a .pdf suffix will be assumed for pdflatex; or what has been declared
% via \DeclareGraphicsExtensions.
\caption{The distributions of mathematical weights obtained by SW-based training and HW-aware
training for Ternary weights are illustrated. In the case of the HW-aware training, a visible stratification has
taken place, separating weights near transition boundaries into distinct categories. When ternarization
is applied to HW-aware trained weights, all mathematical weights will be well within a given discrete
weight level, with very little uncertainty regarding which category the weights belong to. }
\label{weight_distrib}
\end{figure}

The top figure of Fig. \ref{weight_approx} illustrates the hardware approximation function $g^3_{HW}$
at an early stage in the iteration; $w_{sc}$ is relatively large, and the steps across discrete
weight levels are smooth. The bottom figure of Fig. \ref{weight_approx} shows the $g^3_{HW}$ at the end
stage of the iteration; the scale parameter $w_{sc}$ is small, and the steps are nearly abrupt. Both plots
also show the derivative of $g^3_{HW}$ w.r.t. the mathematical weight. It is apparent that the derivative
is non-zero only near value transitions. Thus, in the late stages of the iteration, only mathematical
weights near the transition boundary are impacted by the Gradient Descent family of algorithms (or any other gradient-driven algorithm). Mathematical weights are therefore ``forced to choose'' which side of the
transition boundary they need to be on in order to minimize the cost function. The exact value of the mathematical
weights (beyond the choice of side w.r.t. transition boundaries) ultimately does not matter in a discrete
level system. 
The ``choosing'' effect of the algorithm on weight distributions is illustrated in Fig. \ref{weight_distrib}.
A Ternary NN using the function $g^3_{HW}$ of Eqn. \ref{eq:weights} is trained on MNIST, and the weight
distribution of the hidden layer is compared to that obtained by standard FP training.

Two examples are shown: using a small hidden layer (50 neurons, top plot of Fig. \ref{weight_distrib}),
and a larger hidden layer (200 neurons, bottom plot of Fig. \ref{weight_distrib}). In both cases, the effect
of HW-aware training for the ternary weights is apparent: weight distributions show gaps near the level transition
boundaries. At each step in the iteration, weights near the transition boundary experience a strong gradient
(scaled by the derivative of the $g^n_{HW}$ function, $>>1$ near transitions), which pushes them away from
the boundary in whichever direction minimizes the cost function. Thus, after a few iterations of this algorithm,
no weights (or negligibly few) remain near the boundaries.

\begin{figure}[!ht]
\centering
\includegraphics[width=3.5in]{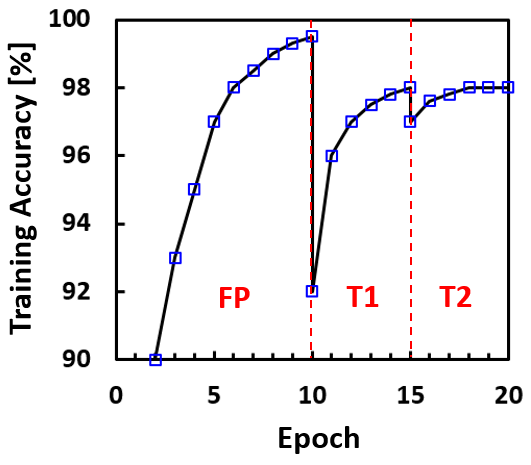}
% where an .eps filename suffix will be assumed under latex, 
% and a .pdf suffix will be assumed for pdflatex; or what has been declared
% via \DeclareGraphicsExtensions.
\caption{The training history of a simple NN with Ternary weights is shown across three iterations
of the HW-aware training algorithm. Initial training (without the HW-aware model) is performed in
the FP set of epochs. The first application of the HW-aware model starts with the T1 set of epochs, 
and initially shows a significant drop in training accuracy, induced by the change in model. The
weights are refined in the next set of epochs, and the final version of the HW-model is applied in T2.
The final training accuracy on the best approximation of the HW nearly matches the FP-based accuracy.}
\label{Training}
\end{figure}

The HW-aware training process is usually completed in two or three iterations. A typical training curve is shown
in Fig. \ref{Training}. The training accuracy is shown vs. Epoch, for a two-iteration training
process. The ``zeroth'' iteration, labeled ``FP'' in Fig. \ref{Training} is just the standard SGD with
FP-based weights and activations. After 10 epochs, the FP iteration reaches \~ 99\% training accuracy.
The ``zeroth'' iteration is terminated, and the next iteration begins. A new neural network is created
using a Ternary HW model for the weights ($g^3_{HW}$), with the initial values for the weights obtained
from the final iteration of the FP network. The Ternary step parameter $\Delta$ is set to 0.45 
(a hyperparameter), while the $w_{sc}$ parameter is set to 0.05. This results in a smooth Ternary discretization
that produces a significantly different model and cost than the FP case. This can be seen
by the abrupt drop in training accuracy at the start of epoch 11. The same set of (mathematical) weights that produced
\~ 99\% training accuracy with the FP model now only produces \~ 91\% with the approximate HW model (T1). However, after additional training with
the initial ternary model T1, the training accuracy is increased to \~ 98\%. The weight transition parameter $w_{sc}$
of 0.05 is not quite small enough to adequately model discrete levels, so an additional iteration is performed with
$w_{sc}$ set to 0.005. This is shown as the ``T2'' step, starting with epoch 15. Due to the change of the model
to a better approximation of the HW, there is another drop in the inference accuracy, which now equals
97\%. A few epochs of additional training with the final HW model push the inference accuracy up to 98\%. 
The transition scale $w_{sc}$ of 0.005 is sufficiently abrupt that further reductions of the parameter make no
difference (as would be shown by an additional iteration, T3, not included here), so the the inference accuracy
achieved at epoch 20 is the final training accuracy with a realistic HW model (in this case, and ideal Ternary).
The neural net evaluated on the accurate HW model has only a 1\% loss in inference accuracy (training) relative
to the FP-based one. In Sec. \ref{sec:results}, the test (not training) accuracy is investigated
for several HW and network examples.

\noindent The optimization of T1 is complete
after 5 epochs (at epoch 15), and T2 begins. As in the case of T1, a new neural network model is created with
a new HW model for the weights (T2), using the same $\Delta$ as T1, but a further decreased  $w_{sc}$ (now set
to 0.005; essentially abrupt). The degradation in training accuracy from T1 to T2 is small, since even T1
is a reasonably good approximation of the final HW model. A few additional epochs of training with T2 recover
the final training accuracy to \~ 98\%. Thus, the final training accuracy on a model that represents an accurate
representation of true Ternary is nearly the same as the training accuracy in FP. It is also evident that absent
training with the $g^3_{HW}$ functions, direct quantization would have resulted in more than 8\% accuracy loss
(the difference between the final FP and initial T1 accuracy). This will be examined further in the context
of test accuracy in the next section.

\FloatBarrier
\section{Results and Analysis}
\label{sec:results}

\begin{figure}[!ht]
\centering
\includegraphics[width=3.5in]{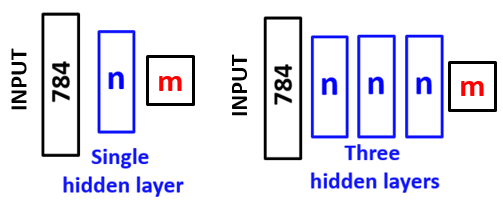}
% where an .eps filename suffix will be assumed under latex, 
% and a .pdf suffix will be assumed for pdflatex; or what has been declared
% via \DeclareGraphicsExtensions.
\caption{The benchmark neural nets used for the analysis are shown. In each case, a neural net
suitable for MNIST (m=10 output classes) or EMNIST (m=47 output classes) is used, 
either with a single hidden layer, or with three identically-sized
hidden layers. In each case, the size of the hidden layer(s) is varied for benchmarking purposes.}
\label{benchmarks}
\end{figure}

\begin{figure}[!ht]
\centering
\includegraphics[width=3.5in]{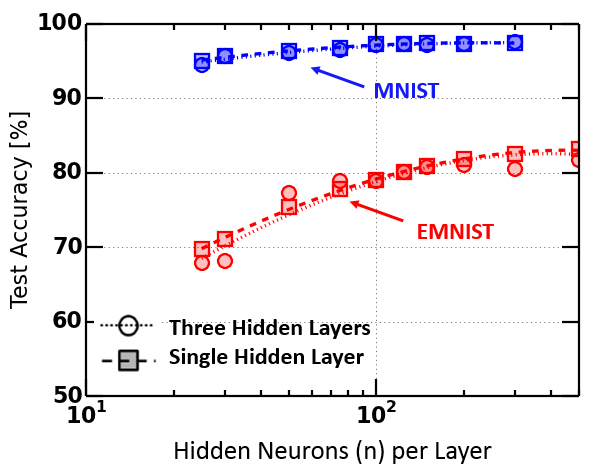}
% where an .eps filename suffix will be assumed under latex, 
% and a .pdf suffix will be assumed for pdflatex; or what has been declared
% via \DeclareGraphicsExtensions.
\caption{The SW-based test accuracies obtained using the benchmark MLPs are illustrated. While the
simple MNIST benchmark easily approaches the 99\% mark, EMNIST is considerably more challenging, and
peaks near 85\% test accuracy. Both figures are consistent with a wide body of literature on MLPs.}
\label{benchmarks_acc}
\end{figure}

The Hardware-aware training algorithm Alg. \ref{algorithm} is next applied to the various architectures
of section \ref{sec:arch}. In each case, the appropriate hardware model is described, and training using
Alg. \ref{algorithm} is performed. The results are benchmarked using MNIST and EMNIST \cite{EMNIST}, with several different
network sizes and two topologies (Fig. \ref{benchmarks}). Two different topologies are used: a single hidden layer,
and three identically-sized hidden layers sandwiched between fixed input and output layers. The size of the hidden
layers is varied, and the inference accuracy on each test set is noted. The obtained SW-based test accuracies
are consistent with expectation based on literature for MLPs \cite{jeds, SYu2, EMNIST}, with MNIST and EMNIST yielding test 
accuracies in the 98\% and 85\% range, respectively (as shown in Fig. \ref{benchmarks_acc}).

\subsection{Binary XNOR}
\begin{figure}[!ht]
\centering
\includegraphics[width=3.75in]{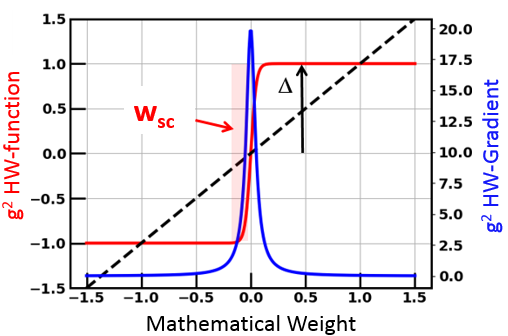}
% where an .eps filename suffix will be assumed under latex, 
% and a .pdf suffix will be assumed for pdflatex; or what has been declared
% via \DeclareGraphicsExtensions.
\caption{The hardware description function for a pure binary weight is shown. Only two weight
levels are available, but no asymmetry is assumed.  }
\label{binary_weight}
\end{figure}

As described in section \ref{sec:xnor}, the particular choice of XNOR circuit results in binary weights
and activations, with no significant distortions of weights due to hardware non-idealities.  This is
a consequence of the dynamic logic implementation of the signal-weight product operation, which results
in either 0 or V$_{DD}$ if the evaluation time is sufficiently long. The accumulation (summation) operation
likewise does not have any weight-induced distortions, although it does, of course, suffer from errors
due to FET Vt-related process variability. Similar considerations
would have resulted had SRAM-based weights been used instead.
Ignoring the variability aspect in this analysis, the weight model
can be described as follows:
\begin{equation}
g^2_{HW} = \Delta\ tanh \bigg( \frac{w}{w_{sc}} \bigg)
\label{eq:binary_weight}
\end{equation}

\noindent where the $2$ superscript denotes the binary weight, $\Delta$ is the weight magnitude
scale parameter (hyperparameter for training), and $w_{sc}$ is the ``hardware realism'' parameter
which determines the transition scale between the ``-1'' and ``1'' states. Note that there is no 
explicit bias term shown in Eqn. \ref{eq:binary_weight}. Instead, the signal vector $X$ is augmented
by one component that is always set to ``1'' (or the appropriate maximum vector value for the given
neural network). The bias is therefore subject to the same binarization as the standard weights.
For hardware implementations where this is not the case, a separate $g$ function could be used
to describe the bias. The approach of augmenting the X-vector for bias is used for all examples
in this paper. The binary weight and
its derivative are illustrated in Fig. \ref{binary_weight}.  

\begin{figure}[!ht]
\centering
\includegraphics[width=3.5in]{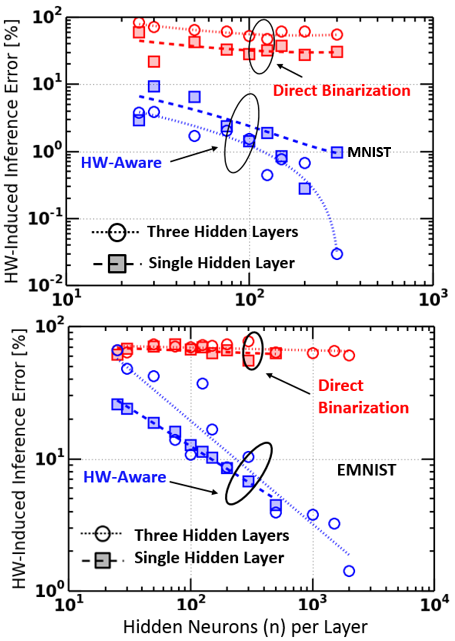}
% where an .eps filename suffix will be assumed under latex, 
% and a .pdf suffix will be assumed for pdflatex; or what has been declared
% via \DeclareGraphicsExtensions.
\caption{The error induced by binarizing SW-trained weights is shown for direct binarization
and using the HW-aware algorithm. The error is defined as the difference of the SW-based and Binary-XNOR inference accuracy. 
Direct binarization exhibits very poor performance with MNIST, with the binarization error never less than 30\%. The HW-aware
algorithm has significantly smaller error for all layer sizes, with the error falling into the 1\% range for
layers of 100 neurons or more. Direct binarization on EMNIST shows extremely poor results, barely better
than random classification. The situation is significantly improved with HW-aware training, with 
errors in the 1\% range possible.}
\label{binary_accuracy}
\end{figure}

Applying the XNOR weight model of Eqn. \ref{eq:binary_weight} and Fig. \ref{binary_weight}
to a pair of simple benchmark networks shows that near-SW level accuracy is achievable. This is 
in contrast to direct binarization of SW-weights, which is shown to result in a significant loss
of inference accuracy (Fig. \ref{binary_accuracy}). 

The direct binarization approach is not able to do better than 30\% error on MNIST w.r.t. the SW implementation,
in spite of a certain degree of optimization of the binarization procedure itself (the magnitude of the
weights is scaled to maximize validation accuracy separately for each network). The situation is
even more dire with EMNIST: the obtained results are barely better than a random network. Direct binarization acts
as a non-linear error amplifier for weights near the weight transition boundary. Small uncertainties
for FP-based weights (from software training) which have negligible impact on inference accuracy
become dramatically amplified by binarization. As indicated in Fig. \ref{binary_accuracy}, this problem is solved by HW-aware training, in which weights
are iteratively pushed to either side of the transition boundary in a way that maximizes inference accuracy.
As seen in Fig. \ref{binary_accuracy}, HW-aware training reduces the error to the 1\% range or less, for 
both MNIST and EMNIST. 
In addition to assigning weights to the optimal side of transition boundaries, the HW-aware algorithm
is also able to re-purpose unused weights to improve inference accuracy. As can be seen from weight distributions
shown in Fig. \ref{weight_distrib}, many (approximately 90\%) of the weights are near zero, and not contributing
to inference. These ``unused'' weights can be put to use by a HW-aware trained network consisting of non-ideal
weights (as in the examples in this work). Thus, a larger number of ``primitive'' weights can perform at the
same level as small number of more complex weights. As long as the network consists of a sufficient number
of ``unused'' weights (as is typical), this is accomplished without adding neurons to the network; simply
re-training in a HW-aware manner is sufficient.
While the training algorithms are quite different, a similar conclusion regarding the capability of 
binary networks is reached in \cite{xnornet}.

\FloatBarrier
\subsection{Ternary Crossbar}

The Ternary crossbar approach of Sec. \ref{sec:Crossbar} provides three weight levels (\{$-2\Delta$, 
0, $2 \Delta$\}, with $\Delta$ a hyperparameter)
which are achieved by signed summation of currents through pairs of conductances ($G^+$, $G^-$). 
The negative sign for
the $G^-$ conductances is attained by a current mirror for the entire crossbar column, which sums $G^-$
currents. With an ideal current mirror, the mirrored current is indeed multiplied by $-1$; in practical
implementations of current mirrors, however, some degree of error is to be expected. A systematic error
in the current mirror results in an asymmetry between the $G^+$ and $G^-$ weights. For the purposes
of this example, $G^-$ will be assumed to be the smaller effective conductance due to imperfect mirroring.
\begin{figure}[!ht]
\centering
\includegraphics[width=3.75in]{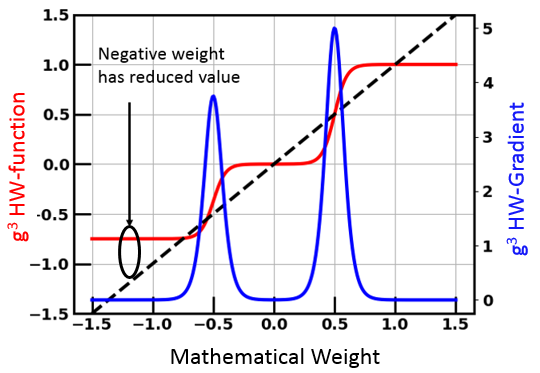}
% where an .eps filename suffix will be assumed under latex, 
% and a .pdf suffix will be assumed for pdflatex; or what has been declared
% via \DeclareGraphicsExtensions.
\caption{The hardware description function for an asymmetric Ternary weight is shown.
The weight function provides three distinct levels with smooth transitions between them.
A high degree of asymmetry was chosen for this example, with the negative weight having
only 75\% of the value of the positive weight. This is likely larger than would be expected
from a practical implementation. }
\label{ternary_weight}
\end{figure}
A suitable hardware model function $g^3_{HW}$ is:

\begin{equation}
g^3_{HW} = 2 \Delta \bigg[ \sigma 
\bigg(\frac{w-\Delta}{w_{sc}}\bigg) \cdot \beta 
+
\sigma \bigg(\frac{w+\Delta}{w_{sc}} \bigg) 
- \beta
\bigg] 
\label{g3}
\end{equation}

\noindent where $\sigma$ is the sigmoid function and $\beta$ defines the degree of asymmetry of the 
($G^+$, $G^-$) pair ($\beta=1$ is perfect symmetry, $\beta=0$ implies $G^-=0$). With $\beta=0.75$,
the behavior of the hardware model is illustrated in Fig. \ref{ternary_weight}.

\begin{figure}[!ht]
\centering
\includegraphics[width=3.5in]{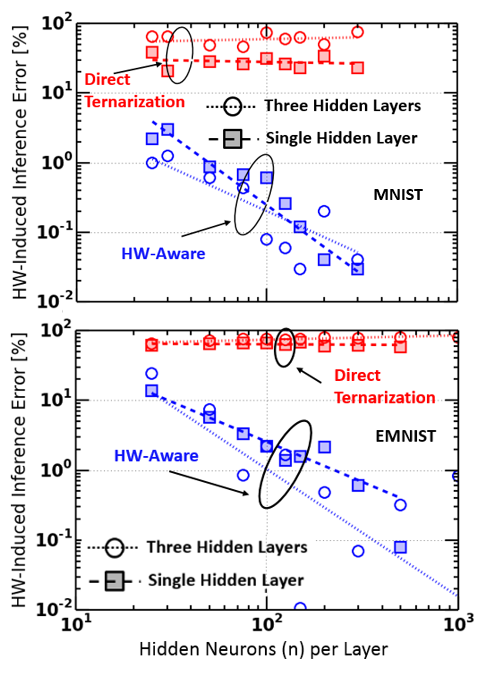}
% where an .eps filename suffix will be assumed under latex, 
% and a .pdf suffix will be assumed for pdflatex; or what has been declared
% via \DeclareGraphicsExtensions.
\caption{The error induced by applying SW-trained weights to a NN with asymmetric ternary weights is compared
to the same using the HW-aware algorithm. The error is defined as the difference of the SW-based and asymmetric Ternary inference accuracy. 
A range of layer sizes is simulated for each topology.
Direct ternarization exhibits very poor performance on both MNIST and EMNIST, with the quantization error never less than 30\%. The HW-aware
algorithm has significantly smaller error for all layer sizes on both benchmarks, with the error falling into the 1\% range for
layers of 100 neurons or more (and approaching 0.1\% for larger networks) }
\label{ternary_accuracy}
\end{figure}

As can be seen in Fig. \ref{ternary_accuracy}, using HW-aware training is of essence for this problem,
although for somewhat different reasons than in the case of the Binary XNOR.
The approach of direct quantization yields very poor performance, with the error relative to the ideal NN never
less than 30\% on MNIST (and considerably worse with EMNIST). Comparing to results in \cite{jeds}, it is clear that most of the error is a result of
the asymmetry, not quantization (unlike in the Binary XNOR case). Direct ternarization in \cite{jeds} with symmetric weights was able to show errors on the order of 10\% (with MNIST), 
indicating the severe penalty imposed by asymmetry in the current example. However, it is also clear that using HW-aware training
almost completely suppresses both the asymmetry and quantization error; the discrepancy between the ideal SW-based
network and the HW-aware trained network is well under 1\% on both benchmarks (MNIST and EMNIST). It should be noted that HW-aware training can only
re-distribute weights across weight categories; there is no change in the hardware values themselves. Thus, the asymmetry
in weight magnitude persists even in the HW (i.e. the current mirrors are still imperfect), but the weight distribution has been optimized
to take this into account. With larger hidden layers, the reduction in the HW-induced error is particularly striking, falling
to less than 0.1\%, even for the more challenging case of EMNIST. This is an illustration of the weight re-purposing nature
of the HW-aware training algorithm. With a large number of weights available, most of the weights are essentially unused
in an FP-trained net (as seen by the weight distributions of Fig. \ref{weight_distrib}, where FP-based weights
cluster around zero). During HW-aware training, these unused weights are re-purposed, helping to overcome their more
primitive nature in HW. The larger the network, the more potential for re-purposing, and consequently the smaller
the resulting HW error.

\FloatBarrier
\subsection{Quinary Crossbar}

\begin{figure}[!ht]
\centering
\includegraphics[width=3.75in]{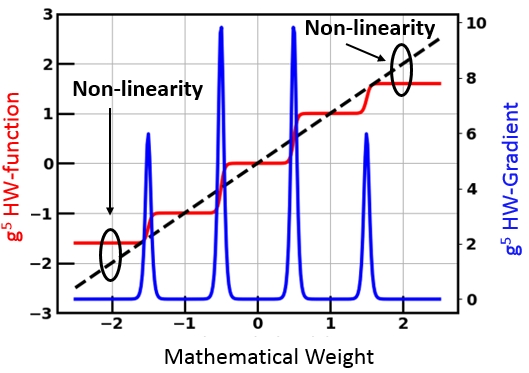}
% where an .eps filename suffix will be assumed under latex, 
% and a .pdf suffix will be assumed for pdflatex; or what has been declared
% via \DeclareGraphicsExtensions.
\caption{The hardware description function for a non-linear Quinary weight is shown. The weight
values deviate from the ideal quantized version, most visibly so for large values of weights.
For this example, the non-linearity is assumed to be symmetric. }
\label{quaternary_weight}
\end{figure}

The Quinary crossbar, with its 2-bit weights, has been shown to be quite accurate when the non-linear 
distortion of the weights is small \cite{jeds, SYu2}. For this example, a large non-linearity will be assumed. The large
non-linearity may result from FET and resistor properties that are less ideal than assumed in \cite{jeds}.
In the context of \cite{jeds}, the most obvious source of non-linearity is the finite FET conductance, which should ideally be much larger than that of the passive series resistor. Depending on the implementation details,
that may not be easily realizable, leading to a larger non-linearity than is described in \cite{jeds}. The
effects of quantization and non-linearity can be handled through HW-aware training, as shown next.
The $g^5_{HW}$ function associated with a non-linear Quinary
weight is shown in Fig. \ref{quaternary_weight}, and expressed mathematically as:
\begin{equation}
\begin{aligned}
g^5_{HW} =  2 \Delta \sum_{k=1,2} \beta_k \bigg[ & \sigma 
\bigg(\frac{w-(2k-1)\Delta}{w_{sc}}\bigg) + \\
           & \sigma \bigg(\frac{w+(2k-1)\Delta}{w_{sc}} \bigg) 
- 1
\bigg] 
\end{aligned}
\label{eq:quinary}
\end{equation}

\noindent where $\beta_k$ is a discrete (k-dependent) function which defines the deviation from
linearity in the HW model of the weight. For a linear model, $\beta_k=1$ for all $k$. More generally, it can be computed using a regression fit to simulation or
measured data which characterizes the HW weight. For the purposes of this example, a strong deviation 
from linearity is assumed, sufficient to significantly degrade inference performance
in the case of direct quantization (since the latter was known to be good in the absence of non-linearity
from \cite{jeds, SYu2}.).

The performance of the Quinary weight with and without HW-aware training is shown in 
Fig. \ref{quaternary_weight}. The non-linearity degrades the expected high accuracy of the 
directly-quantized neural net. As discussed in \cite{jeds}, the expected inference error of the
Quinary weight is as low as 1\% on this problem (on MNIST). With the non-linearity included, the HW inference
accuracy is at best within 10\% of the SW version (for MNIST, and considerably worse with EMNIST). Using the HW-aware training algorithm with 
the model of Eqn. \ref{eq:quinary} reduces the error to the 0.1\% level for both the MNIST and EMNIST
benchmarks. Much like the case of the asymmetric Ternary weight, the network composed of non-linear Quinary
weights benefits considerably from larger layer sizes. As before, the root cause of this behavior is the
re-purposing of near-zero weights during HW-aware training. This excess of essentially unused weights (with 
FP-training) is a common feature of MLPs.

\begin{figure}[!ht]
\centering
\includegraphics[width=3.5in]{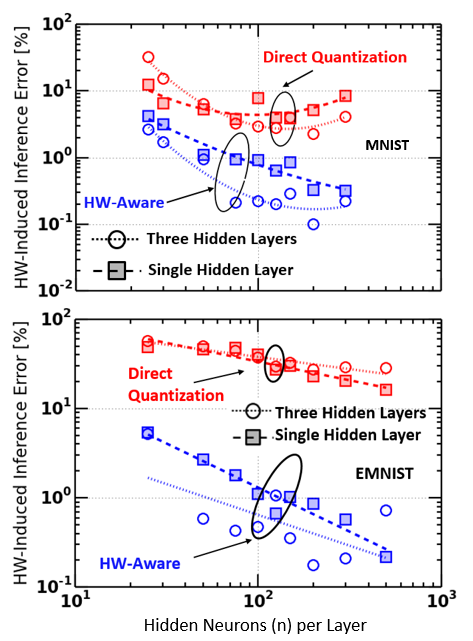}
% where an .eps filename suffix will be assumed under latex, 
% and a .pdf suffix will be assumed for pdflatex; or what has been declared
% via \DeclareGraphicsExtensions.
\caption{The error induced by applying SW-trained weights to a NN with distorted Quinary weights is compared
to the same using the HW-aware algorithm. 
Direct quantization results in moderately poor performance, with the error never below 10\% on EMNIST 
Most of the error is due to non-linear distortion. The HW-aware
algorithm greatly improves on this, with the error falling into the 0.1\% range for both the MNIST
and EMNIST benchmarks. }
\label{quaternary_accuracy}
\end{figure}

\FloatBarrier
\subsection{Other Applications}
\begin{figure}[!ht]
\centering
\includegraphics[width=3.5in]{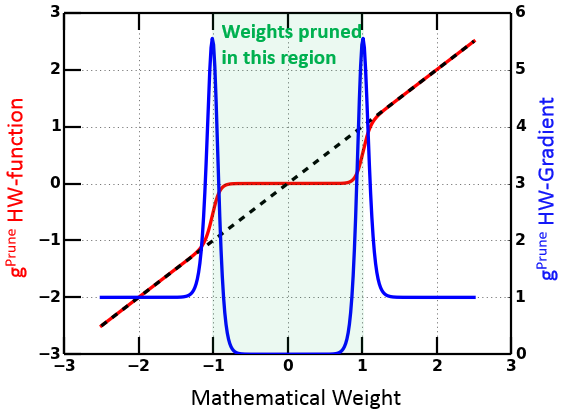}
% where an .eps filename suffix will be assumed under latex, 
% and a .pdf suffix will be assumed for pdflatex; or what has been declared
% via \DeclareGraphicsExtensions.
\caption{The hardware function used for pruning is illustrated. Outside of the pruning window (shown
as \{-1, 1\} in this example) the computed weights are equal to the mathematical weights. Inside the
pruning window (highlighted in green), the weights are set to zero. As in all applications of 
Alg. \ref{algorithm}, at intermediate iterations the hardware function provides smooth transitions
between the pruned and un-pruned regions, eventually becoming abrupt in the final step of the
algorithm. The hardware function itself is shown in red, the derivative is shown in blue.}
\label{pruning}
\end{figure}

% ( weights * (1-self.sigmoid(weights,  -self.x0, self.xsc) +
%		                    1-self.sigmoid(-weights, -self.x0, self.xsc)) )
While all examples shown so far have highlighted the use of Alg. \ref{algorithm} for
cases of non-ideal weights, it is also possible to apply the algorithm to other scenarios in which the
desired behavior of weights and activations does not match that of the originally trained neural net.
As an example, the training algorithm can be used to prune an already trained network in an optimal way.
The pruning results in a much sparser neural net. While the sparsity is not straightforward to exploit in
a crossbar array, it is certainly desirable if the neural net is implemented with a digital ALU. If
the network is made to be sufficiently sparse, the number of weights may be small enough to fit into a local cache,
thereby avoiding the energy and delay penalty associated with DRAM access of weights. For the sake of this example,
the weights are considered to be implemented in a multi-bit digital fashion, with a sufficient number of
bits to be approximated as a continuum. The goal of pruning is to remove as many of the near-zero weights as possible 
without significantly impacting the inference accuracy of the resulting sparse net. In order to accomplish this,
the Alg. \ref{algorithm} is applied with the hardware function shown
in Fig. \ref{pruning}.

\begin{figure}[!ht]
\centering
\includegraphics[width=3.5in]{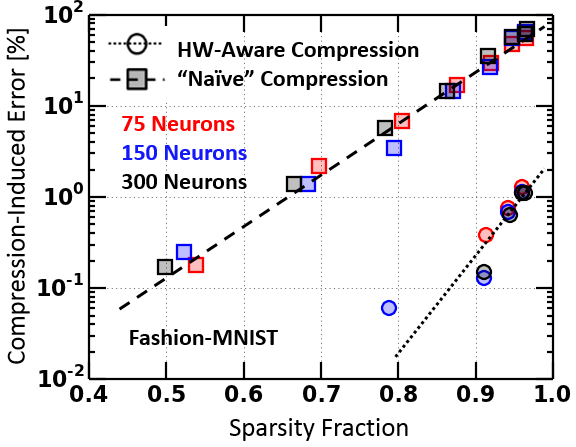}
% where an .eps filename suffix will be assumed under latex, 
% and a .pdf suffix will be assumed for pdflatex; or what has been declared
% via \DeclareGraphicsExtensions.
\caption{The tradeoff between sparsity and inference accuracy is illustrated for the naive and
HW-aware pruning algorithms. Several hidden-layer sizes are shown. It is evident that the naive
algorithm becomes infeasible with even modest weight pruning. The HW-aware algorithm, however,
is shown to suffer only minor reductions in inference accuracy even at a 95\% sparsity
level (i.e. only 5\% of the weights are retained). The effect of layer size on the performance
of each algorithm is seen to be negligible. }
\label{sparsity}
\end{figure}

The pruning hardware function of Fig. \ref{pruning} exhibits two distinct behaviors: in the un-pruned region,
the computed weights are equal to the mathematical weights. In the pruned region, the computed weights
are identically zero. During the training process, the transition between the two regions becomes increasingly
abrupt. The expression for the pruning function of Fig. \ref{pruning} is given as:

\begin{equation}
g_{HW}^{Pr}= w \cdot \bigg[ 1-\sigma \bigg( \frac{w-w_0}{w_{sc}} \bigg) 
                                                + \sigma \bigg( \frac{w+w_0}{w_{sc}} \bigg) \bigg]
\label{eq:pruning}
\end{equation}

\noindent where $\sigma$ is the sigmoid function, $w_0$ is the half-width of the pruning window (assumed 
symmetric in this example), and $w_{sc}$ is the transition scale parameter. The pruning function is
next applied to the same MLPs as in previous examples, but this time using Fashion-MNIST 
\cite{fashion} as the test case. Fashion-MNIST is used for this example since it results in denser
synaptic matrices than either MNIST or EMNIST. Strictly speaking, the matrices for all MLPs are dense; in this
context, ``denser'' simply means having a larger number of weights which are too 
large to be approximated as zero.
It is therefore a more interesting case for pruning. 
The results of pruning are shown in Fig. \ref{sparsity}. The size of the pruning window was varied to
produce a range of sparsities (defined here as the fraction of non-zero values in the synaptic matrices).
Two different approaches to pruning are illustrated: 1. The ``naive'' approach, in which all weights
inside the pruning window are simply set to zero. 2. Alg. \ref{algorithm}
using the HW-function of Eqn. \ref{eq:pruning}. Several sizes of the hidden layers are used, in order
to test the effect of the matrix sizes on the efficacy of the pruning approach. 

The effect of the two pruning approaches is illustrated in Fig. \ref{sparsity}. As the size
of the pruning window is increased, the number of eliminated weights increases, and inference
accuracy is reduced. However, while the naive algorithm shows a steep degradation of inference
accuracy with increasing sparsity, the HW-aware pruning algorithm enables high sparsity levels
with only a minimal impact to accuracy. Unlike the naive algorithm, the HW-aware pruning is free
to re-distribute weights for each size of the pruning window. As was the case in other HW-aware
training examples of this paper, only a very small number of weights are actually moved across
transition boundaries. However, the choice of which weights to modify is guided by cost function optimization; this results in superior performance to simply eliminating ``small'' weights.

\FloatBarrier
\section{Summary and Conclusion}
\label{sec:conclusion}

Some of the challenges of off-line training of neural nets with weight transfer to low-precision hardware
were demonstrated on several examples. The challenges of low bit-width, weight asymmetry, and non-linear
distortion were highlighted as potential sources of loss of inference accuracy. The issue is of
particular importance for edge implementations, where the area savings of low-precision weights and
off-chip learning are essential for improved PPA.
In order to combat the mismatch problem between the models used for off-line training and edge inference,
a new Hardware-Aware training algorithm has been proposed. The proposed algorithm is designed to be general;
it is useable with any model of the hardware, including the aforementioned non-idealities such as weight
asymmetry, non-linear distortion, and quantization. The latter is particularly challenging for training,
since it results in vanishing gradients due to the discrete levels available for weights. The proposed
algorithm handles discontinuities by applying standard training techniques to a sequence of models
approximating discontinuous hardware. At each step in the approximation, the model is continuously differentiable,
but introduces increasingly steep transitions between discontinuous levels. Weights near level-transition
boundaries are pushed by the optimizer to either side, in a manner which minimizes the overall cost function.
As a result, the non-linear amplification of optimization error, which results from direct quantization
of SW-optimized weights, is eliminated. The algorithm has several features which make it attractive
for neural network developers:
\begin{itemize}
\item{It is an incremental algorithm that modifies existing, FP-trained neural nets. 
There is no need to re-train a complex network from scratch.}
\item{It is general w.r.t. the description of the hardware. Any HW model, be it continuous
or not, can be used with the algorithm. The only requirement is an analytical description
of weights, activations, and biases.}
\item{It works with standard neural network training tools; there is no need for a custom
optimization scheme. As such, it is easily used within existing software frameworks.} 
\end{itemize}
The HW-aware training algorithm was applied to several challenging problems for weight transfer, including
pure Binary-XNOR, asymmetric Ternary Crossbars, and non-linearly distorted Quinary Crossbars. Each case
was shown to produce poor results using direct weight transfer, but was improved dramatically using 
HW-aware training. Additional applications of the algorithm were also explored, specifically pruning of 
FP-based networks, with very good initial success. While further exploration of the utility of the proposed algorithm is required, the
presented results suggest it is a promising candidate for straightforward off-line training of edge neural nets.

%\begin{figure}[!h]
%\centering
%\includegraphics[width=2.5in]{LongChannelIMEC.png}
%% where an .eps filename suffix will be assumed under latex, 
%% and a .pdf suffix will be assumed for pdflatex; or what has been declared
%% via \DeclareGraphicsExtensions.
%\caption{The room-temperature Id-Vg characteristics of the long-channel In$_{70}$Ga$_{30}$As GAA are illustrated.
%PBE is negligible at this length. The leakage floor is set by BTBT alone.  Symbols are measured data, lines
%are simulation results.}
%\label{LongChannelIMEC}
%\end{figure}

% or


%\section*{Acknowledgment}


\ifCLASSOPTIONcaptionsoff
  \newpage
\fi



%\bibliographystyle{IEEEtran}

\begin{thebibliography}{1}


\bibitem{Strukov1}
F.M.~Bayat, X.~Guo, M.~Klachko, N.~Do, K.~Likharev, D.~Strukov, 
``Model-based high-precision tuning of NOR flash memory cells for analog computing applications'', 
Proceedings of DRC’16, Newark, DE, June 2016.

\bibitem{jeds}
B.~Obradovic, T.~Rakshit, R.~Hatcher, J.A.~Kittl, M.S.~Rodder,
``Multi-Bit Neuromorphic Weight Cell Using Ferroelectric FETs, suitable for SoC Integration'',
IEEE Journal of the Electron Devices Society 2018 vol. 6, DOI: 10.1109/JEDS.2018.2817628


\bibitem{SYu2}
P.~Yao, H.~Wu, B.~Gao, N.~Deng, S.~Yu, H.~Qian, 
``Online training on RRAM based neuromorphic network: Experimental demonstration and operation scheme optimization'',
2017 IEEE Electron Devices Technology and Manufacturing Conference, EDTM 2017 - Proceedings. Institute of Electrical and Electronics Engineers Inc., p. 182-183.

\bibitem{SYu3}
S.B.~Eryilmaz, D.~Kuzum, S.~Yu, H.S.P.~Wong,
``Device and system level design considerations for analog-non-volatile-memory based neuromorphic architectures'', 
Feb 16 2016 Technical Digest - International Electron Devices Meeting, IEDM. Institute of Electrical and Electronics Engineers Inc., Vol. 2016-February, p. 4.1.1-4.1.4.

\bibitem{GBurr1}
S.~Kim, M.~Ishii, S.~Lewis, T.~Perri, M.~BrightSky, W.~Kim, R.~Jordan, G.W.~Burr, N.~Sosa, A.~Ray, J-P~Han, C.~Miller, K.~Hosokawa, C.~Lam,
``NVM neuromorphic core with 64k-cell (256-by-256) phase change memory synaptic array with on-chip neuron circuits for continuous in-situ learning''
IEEE International Electron Devices Meeting (IEDM) 2015.

\bibitem{GBurr2}
G.W.~Burr, R.M.~Shelby, A.~Sebastian, S.~Kim, S.~Kim, S.~Sidler, K.~Virwani, M.~Ishii, P.~Narayanan, A.~Fumarola, L.L.~Sanches, I.~Boybat, M.L.~Gallo, K.~Moon, J.~Woo, H.~Hwang, Y.~Leblebici,
``Neuromorphic computing using non-volatile memory''
Pages 89-124 | Received 31 Aug 2016, Accepted 01 Nov 2016, Published online: 04 Dec 2016.

\bibitem{burr_nature}
S.~Ambrogio, P.~Narayanan, H.~Tsai, R.M.~Shelby, I.~Boybat, C.~Nolfo, S.~Sidler, M.~Giordano, M.~Bodini, N.C.P.~Farinha, B.~Killeen, C.~Cheng, Y.~Jaoudi, G.W.~Burr,
``Equivalent-accuracy accelerated neural-network training using analogue memory'',
Nature vol. 558, pages 60–67 (2018).

\bibitem{EMNIST}
G.~Cohen, S.~Afshar, J.~Tapson, and A.~Schaik
``EMNIST: an extension of MNIST to handwritten letters''
, arXiv:1702.05373.


\bibitem{xnornet}
M.~Rastegari, V.~Ordonez, J.~Redmon, A.~Farhadi
``XNOR-Net: ImageNet Classification Using Binary Convolutional Neural Networks''
, arXiv:1603.05279.

\bibitem{gxnornet}
L.~Deng, P.~Jiao, J.~Pei, Z.~Wu, G.~Li
``GXNOR-Net: Training deep neural networks with ternary weights and activations without full-precision memory under a unified discretization framework'', arXiv:1705.09283.

\bibitem{ternary_weight_networks}
F.~Li, B.~Zhang
``Ternary weight networks'' ,
arXiv:1605.04711v2.

\bibitem{dorefa}
S.~Zhou, Y.~Wu, Z.~Ni, X.~Zhou, H.~Wen, Y.~Zou
``DoReFa-Net: Training Low Bitwidth Convolutional Neural Networks with Low Bitwidth Gradients'',
arXiv:1606.06160v3.

\bibitem{fashion}
H.~Xiao, K.~Rasul, R.~Vollgraf,
``Fashion-MNIST: a Novel Image Dataset for Benchmarking Machine Learning Algorithms'',
arXiv:1708.07747.


\end{thebibliography}
\FloatBarrier

\end{document}